\documentclass{svproc}
\usepackage{cite}
\usepackage{amsmath,amssymb,amsfonts}
\usepackage{algorithmic}
\usepackage{graphicx}
\usepackage{textcomp}
\usepackage{xcolor}
\DeclareUnicodeCharacter{2212}{-}
\usepackage[utf8]{inputenc}
\usepackage{array}
\usepackage{wrapfig}
\usepackage{multirow}
\usepackage{tabularx}
\usepackage{url}

\begin{document}
\mainmatter % start of a contribution
\title{Analysis of the Cost of Varying Levels of User Perceived Quality for Internet Access}
\titlerunning{Analysis of the Cost} % abbreviated title (for running head)
% also used for the TOC unless
% \toctitle is used
%
\author{Ali Adib Arnab\inst{1} \and John Schormans\inst{2}\and
Sheikh Md. Razibulhasan Raj\inst{1} \and \\Nafi Ahmad \inst{2}}
\authorrunning{Ali Adib Arnab et al.} % abbreviated author list (for running head)
\tocauthor{Ali Adib Arnab, John Schormans, Sheikh Md. Razibulhasan Raj, and Nafi Ahmad}
\institute{University of Global Village, Barisal, Bangladesh,\\
\email{adib9877@yahoo.com},
\and
Queen Mary University of London,London, UK}
\maketitle % typeset the title of the contribution
\begin{abstract}
Quality of Service (QoS) metrics deal with network
quantities, e.g. latency and loss, whereas Quality of Experience
(QoE) provides a proxy metric for end-user experience. Many
papers in the literature have proposed mappings between various
QoS metrics and QoE. This paper goes further in providing
analysis for QoE versus bandwidth cost. We measure QoE using
the widely accepted Mean Opinion Score (MOS) rating. Our
results naturally show that increasing bandwidth increases MOS.
However, we extend this understanding by providing analysis for
internet access scenarios, using TCP, and varying the number
of TCP sources multiplexed together. For these target scenarios
our analysis indicates what MOS increase you get by further
expenditure on bandwidth. We anticipate that this will be of
considerable value to commercial organizations responsible for
bandwidth purchase and allocation.
\end{abstract}
\section{Introduction}
Quality of Experience (QoE) has a significant but complex relationship with Quality of Service
(QOS) and its underlying factors. Mean Opinion Score (MOS) ranges from 1 to 5 (Bad to
Excellent) and represents QoE. Although considerable work has been carried out, in this paper
we have considered QOE within a budget with numerous metrics such as PLP, bandwidth, and
round trip time, TCP sending rate factor, packet buffer lengths and packet bottleneck capacity.
From the curve fitting, we obtained an analytical expression for bandwidth and bandwidth cost.
The goodness of fit is obtained from SSE, R Square, Adjacent R Square and RMSE.
Consequently, we found one equation with variable MOS and bandwidth and another one with
variable bandwidth and bandwidth cost. The analysis has been performed multiple times for
varying the number of TCP sources.
\subsection{Objectives}
The major objective of this research is to identify the mathematical relationship between MoS
and bandwidth cost.
\section{Related work}
QoS has major implications from a policy standpoint [9]. At the present time, users acquire
Internet service according to a fixed price. A simple monthly subscription fee or price according
to the connection time are common. Dial-up modem, Ethernet, cable Ethernet, or DSL are
utilized to provide bandwidth. There may be no bandwidth guarantee or only statistical
bandwidth guarantees – either one can be offered. Conventional Internet applications mainly
utilize TCP as their transport layer protocol. Packet loss is used as a sign of congestion. \\As
soon as the sender experiences network congestion, TCP responds quickly by reducing its
transmission rate multiplicatively. Consequently, the extent of congestion in the network state
partially determines the transmission rate of the TCP sender.\\ The UK is ranked third in terms
of the services and metrics included in this analysis followed by France and Germany [13]. The
UK has occupied third place in a table of the weighted average and lowest available basket
prices across the services which are covered in analyzing bandwidth cost at the varying level.
The US is the most expensive regarding both average and lowest available prices across all the
services while France is the least expensive. The UK’s highest ranking is acquired by the mobile
phone service price, triple-play bundle price, in which category the UK is ranked second
followed by France. The UK is placed fifth concerning the price required to receive fixed voice
services and this is considered the lowest ranking of the UK [14].
\subsection{Abbreviations and Acronyms}\label{AA}
\subsubsection{Mean Opinion Score and its indicator}
Mean Opinion Score is broadly known as MOS, and has become a widespread perceived media
quality indicator [12].
MOS (also superfluously known as the MOS score), is one way of classifying the characteristic
of a phone call. This score is set out of a five-point scale, shown in Table 1 below:
\begin{table}[h!]
\caption {MOS Score vs Performance} \label{tab:title}
\centering
\begin{tabular}{||c c||}
\hline
MOS Score&
Performance \\ [0.5ex]
\hline\hline
5&
Excellent \\
4&
Good \\
3&
Fair \\
2&
Poor \\
1&
Bad \\ [1ex]
\hline
\end{tabular}
\end{table}
MOS of 4.0 or higher is toll-quality. Once within the building, enterprise voice quality patrons
generally expect constant quality while employing their telephone [08].
\subsubsection{Determining PLP and MOS with constant and variable buffer length}
Our first requirement is to determine PLP and MOS for different numbers of TCP sources, round
trip time, bottleneck capacity in packets and buffer length.We take buffer length values from 10
to 1000 [10]
\subsubsection{Plotting of relationship between MOS and bandwidth with constant and variable
buffer length}
We obtain a MOS vs bandwidth graph for different different settings of buffer lengths.
\subsubsection{bandwidth and bandwidth cost relationship for different parameters}
Different countries pays different amounts of money for internet access. It largely depends on
the internet providers, facilities, internet access availability of networking telecommunication
product, infrastructure etc. For example someone in North America may not be paying the same
as someone in Africa. We considered values for UK which will give us range of bandwidth and
estimated average cost for that bandwidth.
\subsubsection{Finding a formula for bandwidth and bandwidth cost}
From our analysis we obtained graph for bandwidth and bandwidth cost. Our goal is to relate
MOS score and bandwidth cost analytically.
\subsubsection{Evaluate how much the bandwidth costs to provide a target MOS level}
The next step was to use existing formulas for PLP and MOS again. Initially we used bottleneck
capacity as one of the parameters to determine PLP. We obtained a formula for MOS and
bandwidth cost and plotted various bandwidth cost value and evaluated MOS scores against
those values. Also, we obtained a similar curve which we previously obtained from MOS vs
bandwidth.
\subsubsection{Varying number of TCP sources}
If we want to differentiate for big organizations and small family houses we need to modify our
formula for MOS by changing number of TCP sources. In big organizations, the number of TCP
sources will be large and for small household it can be much smaller. We modify the number of
TCP sources in our initial formula for MOS. We obtain one graph of MOS vs bandwidth cost for
a larger number of TCP sources and one graph of MOS vs bandwidth cost for smaller number of
TCP sources.
\section{Method}
\subsection{Determining Parameters Value for Experimental Analysis}
An existing formula for PLP in TCP has been used to determine Mean Opinion Score (MOS)
[17]. MOS was plotted against bandwidth and bandwidth has been plotted against cost. The
relationship of MOS and cost has been determined from MOS vs bandwidth and bandwidth vs
cost plots.\\To determine the values of MOS we require PLP values. According to the analytical
expression and performance evaluation of TCP packet loss probability is given by:
\begin{equation} \label{eq1}
\begin{split}
P_i = \frac{32N^2}{3b(m+1)^2(C.RTT+Q)^2}
\end{split}
\end{equation}
\\
N=Number of TCP sources=50\\ C=Bottleneck capacity in packets per second=12500
\\b=number packets acknowledged by an ACK packet=1
\\m=factor by which TCP sending rate is reduced =1/2
\\RTT=Round Trip Time=0.1 second
\\Q= Packet buffer lengths(bits)\\
According to [10] we can obtain MOS from PLPs shown in equation 2 below:
\begin{equation} \label{eq1}
\begin{split}
MOS=1.46* \exp(\exp(−44*PLP))+4.14*\exp(−2.9*PLP)
\end{split}
\end{equation}
Using different buffer length we get values for PLP and MOS (See Table 1). The Q value has
been taken from 10 to 1000. With the same packet buffer length and increasing bandwidth
(bandwidth) 15 Mbps each time, from 15 to 120 we evaluate PLP and MOS.
\begin{table}[h!]
\caption {Q, PLP and MOS value with different Q} \label{tab:title}
\centering
\begin{tabular}{||c c c||}
\hline
Q&
PLP&
MOS \\ [0.5ex]
\hline\hline
10&
7.465263e-01&
4.750997e-01\\
100&
6.503074e-01&
6.280123e-01\\
200&
5.637028e-01&
8.073142e-01\\
400&
4.353297e-01&
1.171447e+00\\
600&
3.462922e-01&
1.516566e+00\\
800&
2.820191e-01&
1.827308e+00\\
1000&
2.341107e-01&
2.099708e+00\\[1ex]
\hline
\end{tabular}
\end{table}
\begin{table}[h!]
\caption {Q, bandwidth, PLP and MOS value with same Q} \label{tab:title}
\centering
\begin{tabular}{||c c c c||}
\hline
Q&
bandwidth&
PLP&
MOS \\ [0.5ex]
\hline\hline
10&
15&
6.503074e-01&
6.280123e-01\\
10&
30&
1.753233e-01&
2.490591e+00\\
10&
45&
7.995852e-02&
3.326486e+00\\
10&
60&
4.556652e-02&
3.824152e+00\\
10&
75&
2.939265e-02&
4.202314e+00\\
10&
90&
2.051913e-02&
4.492741e+00\\
10&
105&
1.513212e-02&
4.712481e+00\\
10&
120&
1.161832e-02&
4.878501e+00\\[1ex]
\hline
\end{tabular}
\end{table}
bandwidth and cost pricing is different worldwide. We can get an estimation of bandwidth vs
cost for United Kingdom which is used as sample data here as a unit of analysis (Ofcom,
2017).\\
bandwidth= [10 30 50 100 200 400 600 800 1000] Mbps\\
Cost= [20 37 40 42 43 45 46 46 46] Dollars\\
\subsection{Determining formula for bandwidth and bandwidth cost by Curve fitting}
To obtain bandwidth vs cost curve, we need a specific formula for bandwidth and cost.
\begin{equation}
\begin{split}
f(x)=ax^b
\end{split}
\end{equation}
Coefficients (with 95\% confidence bounds)\\
a= 27.13 (22.65, 31.61)\\
b= 0.0986 (0.06944, 0.1279)\\
The values of a or b were provided by the curve fitting as the best fit for the graph we got for
bandwidth vs cost. Confidence bounds value are:\\
Goodness of fit:\\
SSE: 38.48\\
R-square 0.9589\\
Adjusted R-square 0.953\\
RMSE 2.345\\
See table IV below:
\begin{table}[h!]
\caption {Goodness of fit parameters for curve fitting} \label{tab:title}
\begin{center}
\begin{tabular}{ | m{3em} | m{3cm}| m{3cm} | }
\hline
Fit name&
Data&
Fit type\\
SSE&
R-square&
DFE\\
Adj R-sq&
RMSE&
coeff \\ [0.000001ex]
\hline
Fit 1&
Cost vs bandwidth&
Power1\\
38.4806&
0.9589&
7\\
0.9530&
2.3446&
2\\[.5ex]
\hline
\end{tabular}
\end{center}
\end{table}
\begin{table}[h!]
\caption {Method and Algorithm for curve fitting} \label{tab:title}
\centering
\begin{tabular}{||c c||}
\hline
Method&
NonlinearleastSquares\\ [0.5ex]
\hline\hline
Robust&
LAR\\
Algorithm&
Trust Region\\
DiffMinChange&
1.0e-8\\
DiffMaxChange&
0.1\\
MaxFunEvals&
600\\
Maxlter&
400\\
TolFun&
1.0e-6\\
TolX&
1.0e-6\\[1ex]
\hline
\end{tabular}
\end{table}
\begin{table}[h!]
\caption {Method and Algorithm for curve fitting} \label{tab:title}
\centering
\begin{tabular}{||c c c c||}
\hline
Coefficient&
StartPoint&
Lower&
Upper\\ [0.5ex]
\hline\hline
a&
9.2708&
-lnf&
lnf\\
b&
0.3019&
-lnf&
lnf\\[1ex]
\hline
\end{tabular}
\end{table}
From Table VI we know the method named in curve fitting is NonlinearleastSquares which is
taken automatically by the characteristics, shape and number of squares present in the curve.
Among Interpolant, Linear fitting, Polynomial, Rational, Sum of shine, Smoothing Spline,
Weibull, Exponential, Gaussian, Fourier we selected Power1 since it gives better visualization
and better goodness of fit prediction. The prediction includes calculation of SSE, R-square,
Adjacent R-square, RMSE, coefficient with confidence bound. The robust option is enabled and
Least Absolute Residuals (LAR) shows more accurate results than Bi Square. LAR mainly
focuses on the contrast between residuals rather than squares and Bi-Square reduces weight of
the sum of the squares which is necessary for our curve’s case.\\
We will take bandwidth as ‘x’ which lies X axis and bandwidth cost as ‘f(x)’ which lies in Y axis.
We can acquire values for ‘a’ and ‘b’ from curve fitting and by implementing the equation (3) we
obtain equation (4) below:
\begin{equation}
Cost=27.13*bandwidth^{0.0986}
\end{equation}
\begin{figure}[htbp]
\centering
\includegraphics[width=12cm]{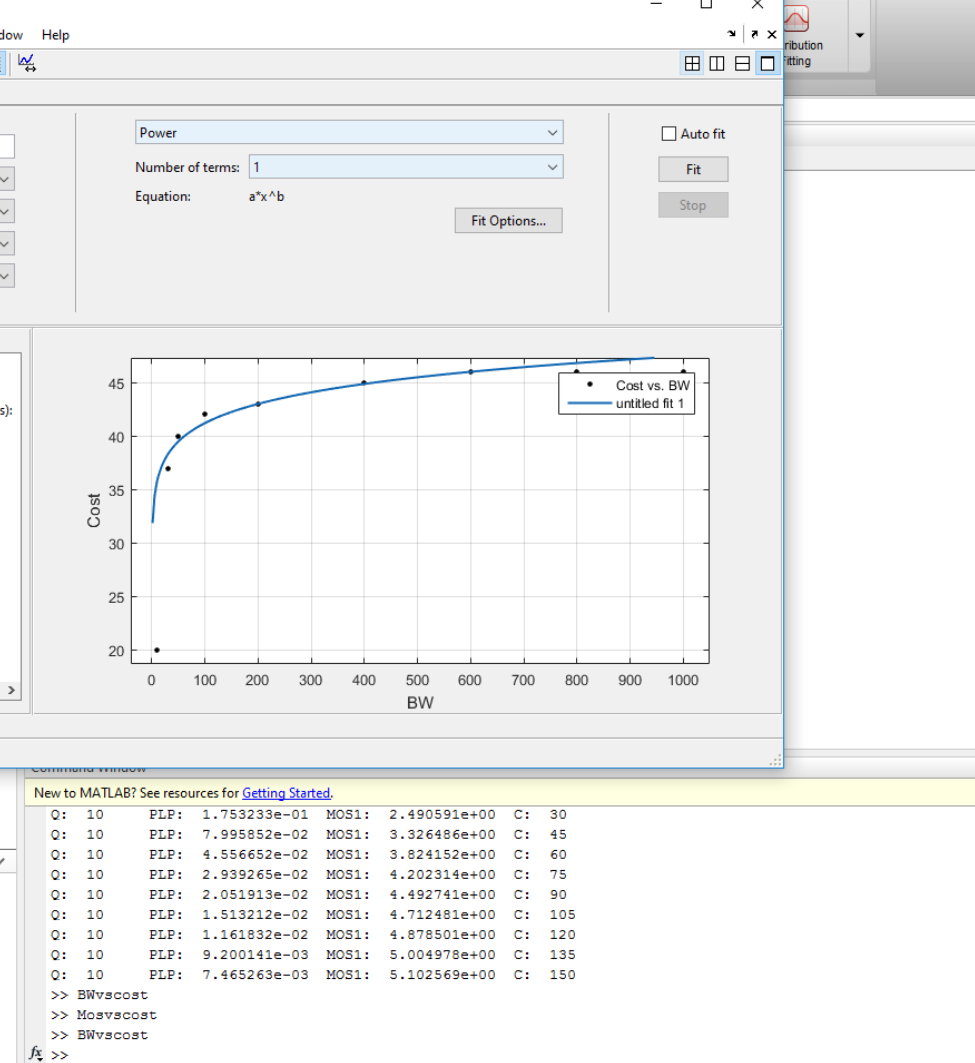}
\caption{Setup for applying Curve fitting formula from Cost vs bandwidth to obtain a
formula for MOS vs bandwidth cost}
\label{fige:asa}
\end{figure}
\subsection{Implementing MOS and cost relationship by eliminating bandwidth from bandwidth
vs cost formula}
By taking different values of cost we can get different values of bandwidth and if we replace
bandwidth from equation (2) we can get different values for cost and MOS.
\begin{figure}[htbp]
\centering
\includegraphics[width=12cm]{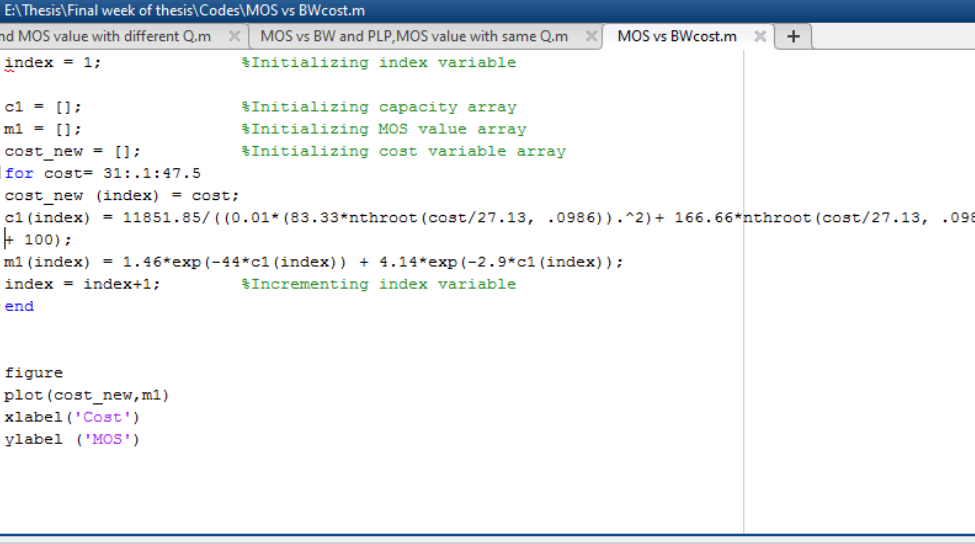}
\caption{MOS vs bandwidth cost relationship}
\label{fige:asa}
\end{figure}
Bottleneck capacity is given by eqn (5) below:
\begin{equation} \label{eq1}
\begin{split}
C = \frac{bandwidth*1000000}{12000}
\end{split}
\end{equation}
\subsubsection{MOS and bandwidth}
From equation (1), (2) and (5) we obtained relationship between MOS and bandwidth which is
equation (6):
\begin{figure}[htbp]
\centering
\includegraphics[width=12cm]{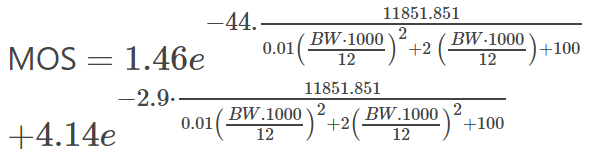}
\label{fige:asa}
\end{figure}
\subsubsection{MOS and bandwidth Cost}
If we put bandwidth cost in the place of bandwidth by help of equation (4) we obtain the
following relationship which is equation (7):
\begin{figure}[htbp]
\centering
\includegraphics[width=12cm]{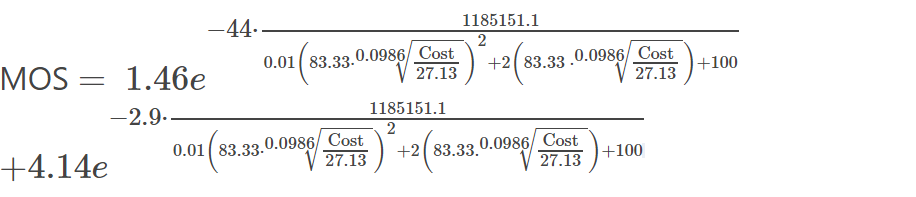}
\label{fige:asa}
\end{figure}
\subsection{MOS and bandwidth Cost}
To evaluate MOS with different numbers of TCP sources we changed the values for N. We took
2 sample values of N 80 and 500. 80 TCP sources mainly represent a small building or
bandwidth use for family purposes. 500 TCP sources represents bigger companies and
organizations.\\
To get MOS and cost relationship, we took a value of N=80 and 500 instead of 50 which
provided us a different PLP and a different MOS. the bandwidth and cost relationship remains
same as before because it is here seen to have nothing to do with the number of TCP sources.
We were able to obtain different MOS and bandwidth formula and different MOS and Cost
formula and get output for different number of TCP sources.\\
Which is denoted by the equation (8):
\begin{figure}[htbp]
\centering
\includegraphics[width=12cm]{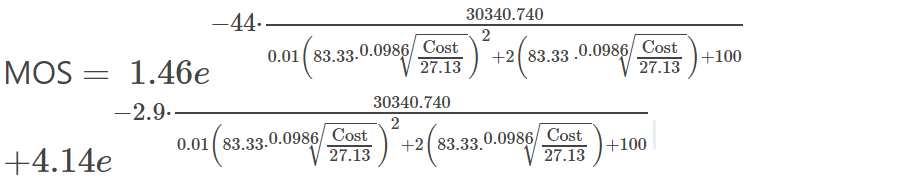}
\label{fige:asa}
\end{figure}
For N=500, MOS and Cost which is equation (9):
\begin{figure}[htbp]
\centering
\includegraphics[width=12cm]{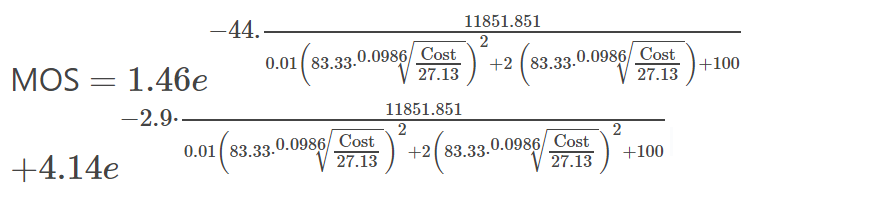}
\label{fige:asa}
\end{figure}
\section{Results and Testing}
\subsection{Plotting MOS vs bandwidth (Bps) with different packet buffer lengths}
\begin{figure}[htbp]
\centering
\includegraphics[width=12cm]{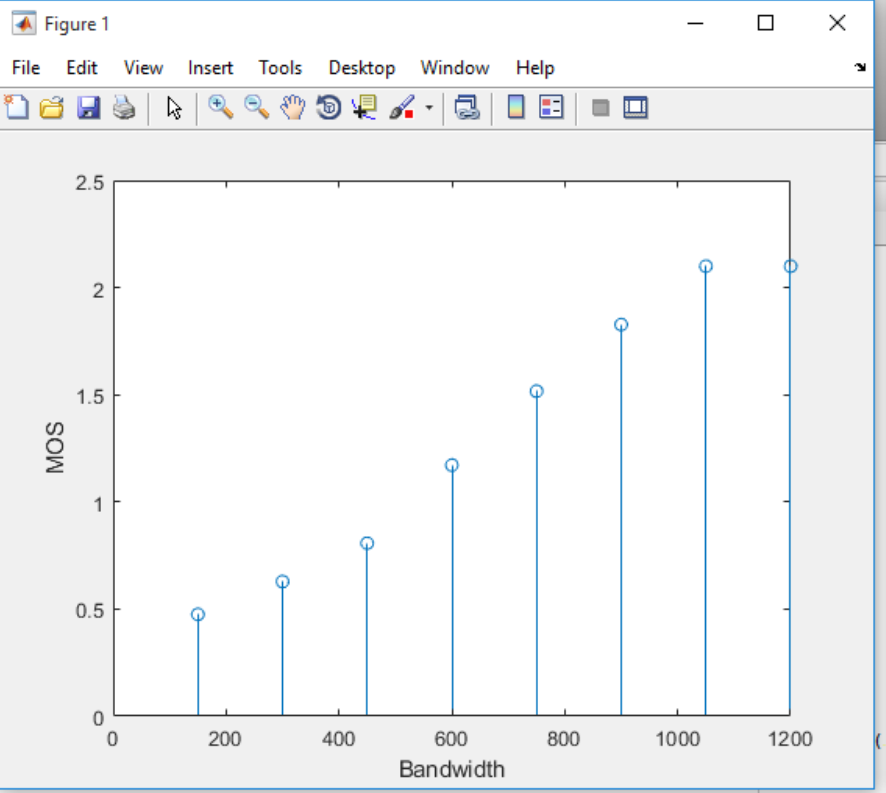}
\caption{MOS vs bandwidth (Bps) with different packet buffer lengths}
\label{fige:asa}
\end{figure}
Our initial formula for PLP provides a relationship between packet loss probability and various
parameters including the number of TCP sources, bottleneck capacity, round trip time, number
of packets acknowledged by an ACK packet, factor by which TCP sending rate is reduced and
packet buffer length. We can determine the PLP by taking sample data for this parameter. The
MATLAB code and formula for PLP is discussed in 5.1. we then calculated MOS from PLP. Fig
4 is MOS vs bandwidth (Mbps) with different packet buffer lengths.
\subsection{Plotting of MOS vs bandwidth (Mbps) with constant packet buffer length Output}
\begin{figure}[htbp]
\centering
\includegraphics[width=12cm]{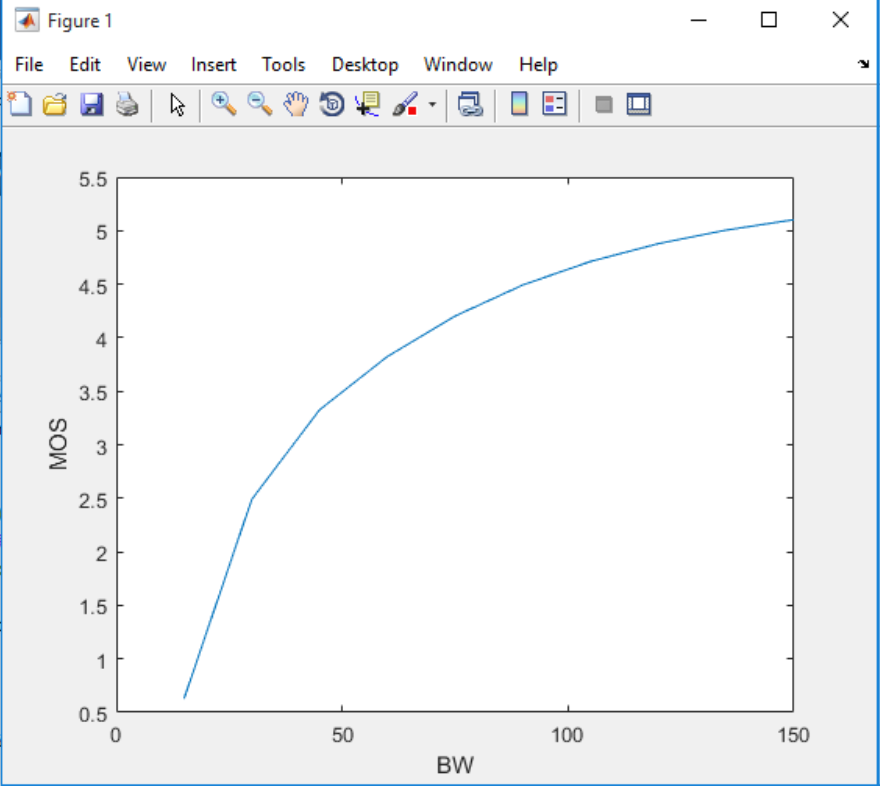}
\caption{MOS vs bandwidth (Mbps) with same packet buffer length}
\label{fige:asa}
\end{figure}
If we keep changing buffer length it is very difficult to evaluate MOS change with the effects of
bandwidth change. So by keeping the same buffer length, Q which is 10, we can obtain a MOS
vs bandwidth curve. So it is quite evident from Fig (4) that when we increase bandwidth, MOS is
also increasing proportionally bandwidth. As a sample when bandwidth is 50 Mbps, MOS is 3.5
(approximately), when bandwidth is 100 Mbps MOS is 4.5 (approximately) and when bandwidth
is 150 Mbps MOS is close to 5.
\subsection{Plotting of bandwidth Vs bandwidth Cost Relationship}
\begin{figure}[htbp]
\centering
\includegraphics[width=12cm]{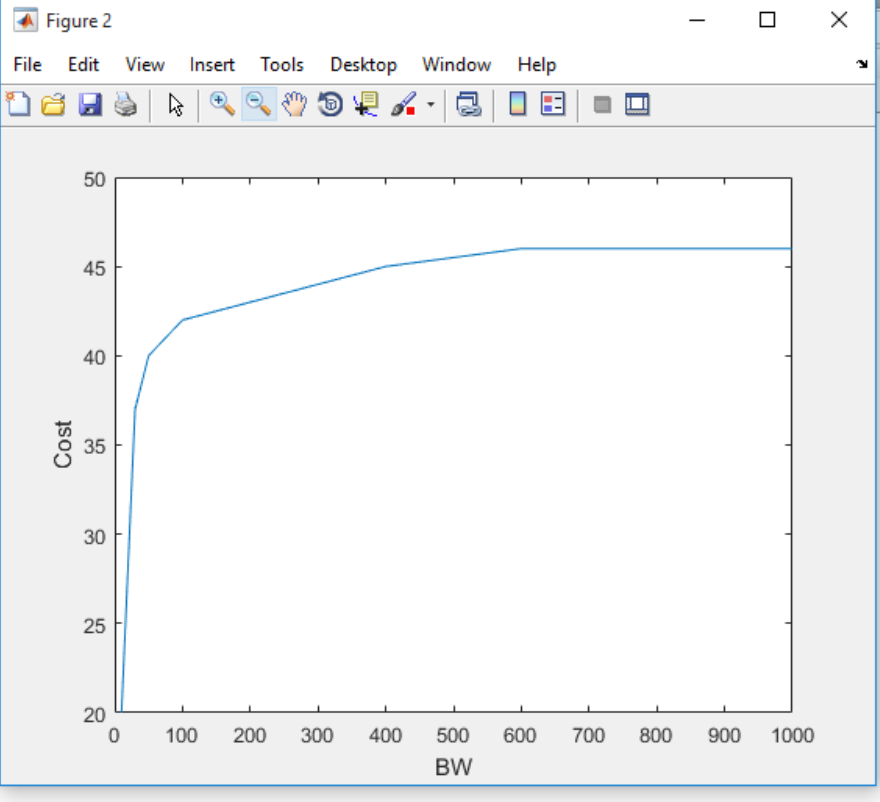}
\caption{bandwidth vs bandwidth cost relationship}
\label{fige:asa}
\end{figure}
We took some different parameters for bandwidth cost in the UK. If we look at the figure (5) we
can see initially the cost increases within a limited range of bandwidth. In this case even for 20
Mbps the bandwidth price is somewhere close to 35£ per month. The rate increases until 100
Mbps, from the graph customer has to pay nearly 43£ for 100 Mbps per month. Beyond 100
Mbps the cost increment is very slow. From 100 Mbps to 1000 Mbps cost only increases from
43 £ to 46£ per month which is incredibly low. So we can draw conclusions about how much it is
worth spending on bandwidth before we run into a law of diminishing returns.
\subsection{Plotting of MOS vs bandwidth cost relationship output}
\begin{figure}[htbp]
\centering
\includegraphics[width=12cm]{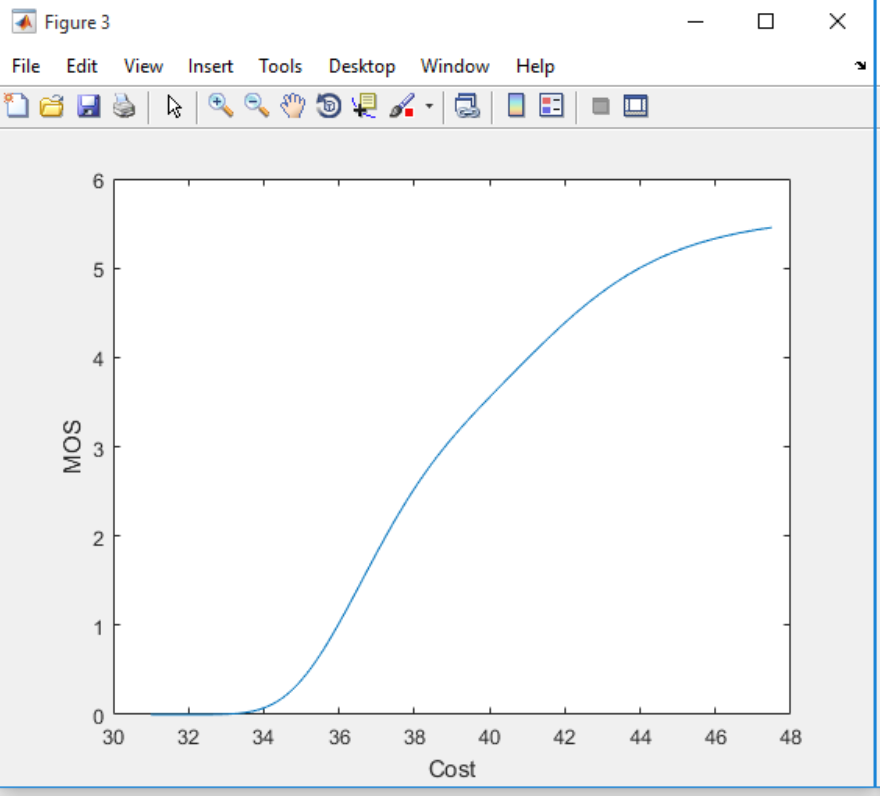}
\caption{MOS vs bandwidth cost relationship}
\label{fige:asa}
\end{figure}
Initially MoS is very low with low cost. That indicates there is a certain amount of money a
customer needs to pay initially to get internet access. A customer paying 38£ can obtain quality
of experience of 2 MoS. Where a customer paying 45£ is receiving far better quality of
experience, MoS is close to 5.\\The experiment makes sense if we compare broadband prices
in the UK. For example, Broadband provider named ‘Now Broadband’ offers 11 Mbps in 18£ per
month in the ‘Brilliant Broadband’ plan. The Same broadband provider is offering 36 Mbps
speed in 24£ per month in the ‘Fab Fibre’ plan. So the bandwidth is increasing more than 3
times while cost is only increasing by 6£ (Cable.co.uk, 2019).
\subsection{Plotting of MOS and bandwidth cost Output with different TCP sources}
\begin{figure}[htbp]
\centering
\includegraphics[width=12cm]{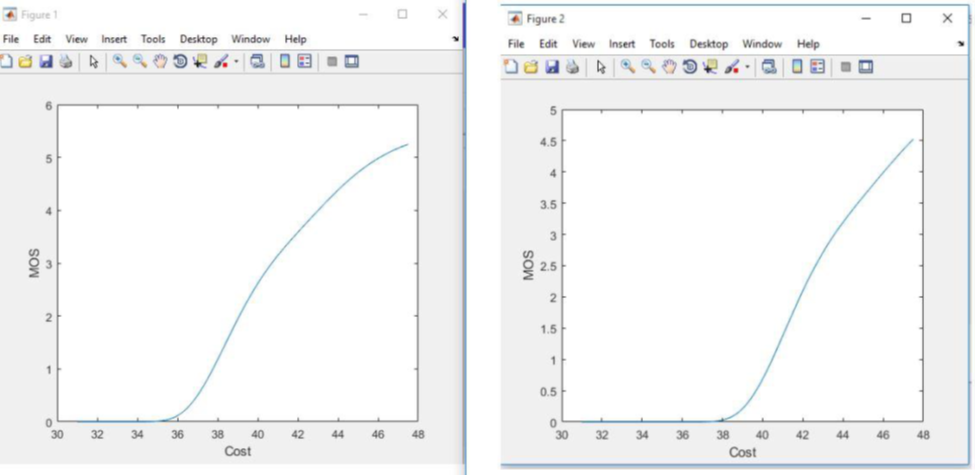}
\caption{MOS vs bandwidth Cost Output with different TCP sources.}
\label{fige:asa}
\end{figure}
Initially we took the number of TCP sources to be 50. If we change the TCP sources we obtain
different outputs and results. There are 2 output graphs in Fig 7. The first one in Fig(7) is
calculated by taking 80 TCP sources and second one is calculated taking 500 TCP sources. If
we take more TCP sources the quality increases more rapidly with the price increase compared
to taking fewer TCP sources. Big organizations usually have TCP sources and so that they have
the luxury of getting better quality of experience within the same cost but after crossing initial
price barrier. In small household fewer TCP sources are most likely used which also
experiences a rise in MOS which provides good quality within a price range but that is less than
that seen in big organizations.
\section{Discussion and Further Work}
Prior work in the field has shown that QoE has a complex relationship with QoS factors like
packet loss probability (PLP), delay or delay jitter. Furthermore, contemporary analyses have
indicated that the relationship between these QoS factors and QoE can significantly vary from
application to application. In this paper we take prior analyses of the relationship between the
key QoS metric of packet loss probability and QoE and target an internet access scenario. We
use this relationship to show how QoE (measured as MOS) varies as more money is spent on
the bandwidth of the internet access link. Our results target two different scenarios – a small
number of multiplexed TCP sources and a relatively large number of multiplexed TCP sources.
We show that increase in MOS is not a linear with increasing spent on bandwidth at all, and
considering these two different we are able to resolve the highly non-linear fashion in which
MOS does increase with spend on bandwidth.

\end{document}